\documentstyle[12pt]{article}
\newcommand{\beq}{\begin{equation}}
\newcommand{\eeq}{\end{equation}}
\newcommand{\beeq}{\begin{eqnarray}}
\newcommand{\eeeq}{\end{eqnarray}}

\begin{document}
\author{V.A. \ Kudryavtsev  \\ 
{\em Petersburg Institute of Nuclear Physics,}\\{\em \ 188350, Gatchina,
Russia.}}
\title{{\bf Spinor representations of the Virasoro and super-Virasoro
algebras for conformal spin to be equal $\frac {1}{k}$ }} \maketitle

\begin{abstract}
It is considered here the possibility of unitary
spinor representations of the Virasoro and super-Virasoro  algebras for
conformal spin to be equal $\frac {1}{k}$; k are integers.
\end{abstract}

  Virasoro group on the two-dimensional plane of complex
  variable z  is defined by the following group of conformal
  transformations :
\beq
 z'^n=\frac{az^n+b}{cz^n+d}; \;\;\;\;\;\;\;\; ad-bc=1;n\in Z
\eeq
  Virasoro generators $L_n$  are determined by transformations:

\beq
 L_n=z^{-n+1}\frac{d}{dz}
\eeq

 They satisfy the usual classical commutator algebra:

\beq
   \left[L_{n},L_{m}\right]=(n-m)L_{n+m}
\eeq

   Quantum mechanical applications of the Virasoro  algebra lead to
   central extensions:

\beq
   \left[L_{n},L_{m}\right]=(n-m)L_{n+m} + c_{mn}
\eeq

  Where
\beq
  c_{mn}=\frac{c}{12} m(m^2-1)\delta_{m,-n}
\eeq

  Unitary conditions for these operators $L_{n}$ are
\beq
   L_{n}^{\dag}=L_{-n}
\eeq

 In physical applications (vertices of string theory) we use
  operator functions  F(z):
\beq
 F(z)=\sum_{n}F_n z^n
\eeq
  which are representations  of this Virasoro group corresponding
 to conformal spin $j$:
\beq
  \left[L_{n},F(z)\right]=(z^{-n+1}\frac{d}{dz}-jn z^{-n})F(z)
\eeq
  Then we should have for any conformal spin j :
\beq
\left[L_{n},F_r\right]=((j-1)n-r)F_{n+r}
\eeq

  Unitary representations of the Virasoro and super-Virasoro
  algebras are found for conformal spins j to be equal $0,1$ and $\frac
  {1}{2}$ \cite {1}.

    For j=1 we have:

\beq
 F_{j=1}(z)=\sum_{n=1}a_n z^n +p + \sum_{n=1}a_{-n} z^{-n}
\eeq
\beq
  \left[a_{n},a_{m}\right]=n\delta_{n,-m}; \;\;\;n>0\;\;\;
a^{\dag}_{n}=a_{-n} ;
\eeq
\beq
L_n=\frac{1}{2}\sum_{m\neq 0}a_{n-m}a_{m}+pa_{n};\;\;\; n\neq 0 ;\;\;\;
L_0=\sum_{m\neq 0}a_{-m}a_{m}+\frac{1}{2}p^2  ;\;\;\;
L_{n}^{\dag}=L_{-n}
\eeq

    For $j=\frac{1}{2}$ we have constructions with components $b_{r}$
(r are half-integers) to be anticommuting:

\beq
 F_{j=\frac{1}{2}}(z)=\sum_{r=\frac{1}{2}}b_r z^r
+\sum_{r=\frac{1}{2}}b_{-r} z^{-r};\;\;\;\;
r=\frac{1}{2};\frac{3}{2};\frac{5}{2}...
\eeq
\beq
\left\{b_{r},b_{s}\right\}=\delta_{r,-s}; \;\;\;
b^{\dag}_{r}=b_{-r} ;
\eeq
\beq
L_n=\frac{1}{2}\sum_{r}(-\frac{n}{2}+r)b_{n-r}b_{r};\;\;\; n\neq 0 ;\;\;\;
L_0=\sum_{r>0}rb_{-r}b_{r};\;\;\;
L_{n}^{\dag}=L_{-n}
\eeq

      Of course we are able to obtain boson representations for
  arbitrary conformal spin as composite ones from j=0
components.  They are exponential vertex operators of
Veneziano type \cite{2}:
\beq
F_{j}(z)= \exp{(-k\sum_{n>0}a_{-n}\frac{z^{-n}}{n})}
\exp{(ikx_0+kp\ln{z})}\exp{(k\sum_{n>0}a_{n}\frac{z^n}{n})};\;\;\;
\eeq
 Here
\beq
\left[p,x_0\right]=\frac{1}{i};\;\;\;j=\frac{k^2}{2}
\eeq

    Now we shall give a spinor consruction  for $j=\frac{1}{4}$.
We use anticommuting spinor components $\psi_r$ (r are
multiples of quarters):
\beq
  \left\{\tilde\psi_{\alpha,r},\psi_{\beta,s}\right\}=
 \delta_{\alpha,\beta}\delta_{r,-s};\;\;\;
\psi_{\alpha,r}^{\dag}= \psi_{\alpha,-r};\;\;
  r=\pm\frac{1}{4};\pm\frac{3}{4};\pm\frac{5}{4}...
\eeq
\beq
\tilde\psi=\psi T_{0};\;\;\;\;\;\;
(T_{0})_{\alpha\beta}=(T_{0})_{\beta\alpha}
\eeq
  Let us build currents $J(z)$ from these spinor components $\psi_r$
and then we introduce the Sugawara-like operators for $L_n$ (compare
\cite{3}):
\beq
 J(z)=
\sum_{n=1}J_n z^n +J_0 + \sum_{n=1}J_{-n} z^{-n};\;\;\;\;
 J_n= \sum_{r}\tilde\psi_{n-r}\Gamma\psi_r;\;\;\;\; \\
\eeq
\beq
 J_0=
\sum_{r}\tilde\psi_{-r}
\tilde\Gamma\psi_r;\;\;\;\;
\tilde\Gamma=\frac{1}{\sqrt{\rho}}\Gamma
\eeq Here $$ J_{-n}=
J_n^{\dag}\;\;\;(T_{0}\Gamma)_{\alpha\beta}=-(T_{0}\Gamma)_{\alpha\beta}
\;\;\;\;\Gamma^2=\rho I;\;\;\;\;    Tr\Gamma^2=1;\;\;\;Tr\Gamma=0
$$
\beq
L_n= \frac{1}{4}\sum_{m}J_{n-m}J_{m};\;\;\;\;n\neq 0 ;\;\;\;\;
L_0= \frac{1}{4}J_{0}^2+\frac{1}{2}\sum_{m>0}J_{-m}J_{m};\;\;\;\;
L_{n}^{\dag}=L_{-n}
\eeq
  $J_n$ satisfy commutation relations:
\beq
  \left[J_{n},J_{m}\right]=2n\delta_{n,-m};\;\;\;\;n>0
\eeq
  $L_n$ satisfy the extended Virasoro  algebra (5).

   J(z) has the conformal spin j to be 1 in relation to $L_n$ (22).
 Now we can obtain the spinor representation $\Psi(z)$ for
$j=\frac{1}{4}$ in relation to $L_n$ (22) acting on vacuum state
$<0|$.  \\
Here $\psi_{\alpha,r}|0>=<0|\psi_{\alpha,-r}=0$ ; $r>0$  \\
\beq <0| \Psi(z)= <0|\sum_{r} \Psi_r z^r
\eeq We take
\beq <0|\Psi_
{\frac{1}{4}}=<0|\psi_{\frac{1}{4}}
\eeq
\beq
<0|\Psi_{\frac{1}{4}}L_0=<0|\psi_{\frac{1}{4}}(\frac{1}{4}J_{0}^2)=
\frac{1}{4}<0|\psi_{\frac{1}{4}}
\eeq
     Other components of $\Psi(z)$ we are able to obtain using
\beq
\left[L_{1},\Psi_r\right]=(-\frac{3}{4}-r)\Psi_{r+1}
\eeq
  So in correspondence with (10),(27) we have
\beq
<0|\Psi_{\frac{5}{4}}=-<0|\left[L_1,\psi_{\frac{1}{4}}\right]=
\frac{1}{2}<0|(\tilde \Gamma \psi_{\frac{1}{4}})J_1
\eeq

 Similary we have
\beq
<0|\Psi_{\frac{9}{4}}=-\frac{1}{2}<0|\left[L_1,\Psi_{\frac{5}{4}}\right]=
<0|(\frac{1}{4}(\tilde \Gamma \psi_{\frac{1}{4}})J_2+
\frac{1}{8}\psi_{\frac{1}{4}}J_1^2)
\eeq
 and
\beq
<0|\Psi_{\frac{13}{4}}=-\frac{1}{3}<0|\left[L_1,\Psi_{\frac{9}{4}}\right]=
<0|(\frac{1}{6}(\tilde \Gamma \psi_{\frac{1}{4}})J_3+
\frac{1}{8}\psi_{\frac{1}{4}}J_1J_2+
\frac{1}{48}(\tilde \Gamma\psi_{\frac{1}{4}})J_1^3)
\eeq
    and finally we have
\beq
<0|\Psi(z)=<0|z^{\frac{1}{4}}((\tilde \Gamma
\psi_{\frac{1}{4}})
\sinh{(\sum_{n>0}J_{n}\frac{z^n}{2n})}+\psi_{\frac{1}{4}}
\cosh{(\sum_{n>0}J_{n}\frac{z^n}{2n})})
\eeq
   It is easy to find the correct correlation function:
\beq
 <0|\Psi(z)
\Psi^{\dag}(1)|0>=\frac{z^{\frac{1}{4}}}{{\rho}\sqrt{(1-z)}}
\eeq

  The required construction of super Virasoro algebra  appears due to
introduction of an additional field $\Phi_{j=\frac{1}{2}}(z)$ and
corresponding supergenerators:  \beq
G_{r}=\frac{1}{\sqrt{2}}\sum_{n}J_n\Phi_{r-n}
\eeq

\beq
\Phi_{j=\frac{1}{2}}(z)=\sum_{r=\frac{1}{2}}\Phi_r z^r
+\sum_{r=\frac{1}{2}}\Phi_{-r} z^{-r};\;\;\;\;
r=\frac{1}{2};\frac{3}{2};\frac{5}{2}...
\eeq
\beq
\left\{\Phi_{r},\Phi_{s}\right\}=\delta_{r,-s}; \;\;\;
\Phi^{\dag}_{r}=\Phi_{-r} ;
\eeq
  So we obtain the necessary super Virasoro algebra:
\beq
  \left\{G_{r},G_{s}\right\}=
2L_{r+s}+\frac{c'}{3}(r^2-1/4)\delta_{r,-s} \eeq \beq
   \left[L_{n},L_{m}\right]=(n-m)L_{n+m}+\frac{c}{12}n(n^2-1)\delta_{n,-m}
\eeq
\beq
   \left[L_{n},G_{r}\right]=(n/2-r)G_{n+r}
\eeq

\beq
L_n= \frac{1}{4}\sum_{m}J_{n-m}J_{m})+
\frac{1}{2}\sum_{r}(-\frac{n}{2}+r)\Phi_{n-r}\Phi_{r};\;\;\;n\neq 0
\eeq
\beq
L_0= \frac{1}{4}J_{0}^2+\frac{1}{2}\sum_{m>0}J_{-m}J_{m}+
\sum_{r>0}r\Phi_{-r}\Phi_{r};\;\;\;
L_{n}^{\dag}=L_{-n}
\eeq
  Generalization of this formalism for $j=\frac{1}{k}$ with any
integer $k>2$ is evident.

\end{document}